\documentclass[floatfix, prb, twocolumn, superscriptaddress, showpacs]{revtex4}

\usepackage{graphicx}

\usepackage{dcolumn}

\usepackage{bm}

\begin{document}

\preprint{\today}
\title{Far-infrared signature of the superconducting gap in intercalated graphite CaC$_6$.}

\author{U.~Nagel}
\email{unagel@kbfi.ee}
\affiliation{National Institute of Chemical Physics and Biophysics, Akadeemia tee 23, 12618 Tallinn, Estonia.}

\author{D.~H{\"u}vonen}
\affiliation{National Institute of Chemical Physics and Biophysics, Akadeemia tee 23, 12618 Tallinn, Estonia.}

\author{E.~Joon}
\affiliation{National Institute of Chemical Physics and Biophysics, Akadeemia tee 23, 12618 Tallinn, Estonia.}

\author{J.\,S.~Kim}
\affiliation{Max-Planck-Institut f\"ur Festk\"orperforschung, Heisenbergstra\ss e 1, D-70569 Stuttgart, Germany}

\author{R.\,K.~Kremer}
\affiliation{Max-Planck-Institut f\"ur Festk\"orperforschung, Heisenbergstra\ss e 1, D-70569 Stuttgart, Germany}

\author{T.~R\~o\~om}
\affiliation{National Institute of Chemical Physics and Biophysics, Akadeemia tee 23, 12618 Tallinn, Estonia.}


\begin{abstract}

Terahertz reflectance spectra of the Ca-intercalated graphite CaC$_6$ reveal a superconducting gap below 11\,K.
The gap signature lacks a sharp onset to full reflectivity at $2\Delta_{0}$, but rather shows a distribution of gap values consistent with an anisotropic gap. 
The experimental data were successfully fitted to the gap distribution obtained from density functional calculations by Sanna \textit{et al.} (Phys. Rev. B75, 020511, 2007).
The temperature dependence of the superconducting gap is characteristic for a BCS type superconductor.

\end{abstract}


\maketitle


%

CaC$_6$ is exceptional in the series of intercalated graphite compounds because of its high superconducting transition temperature, $T_c \approx 11.5$\,K. \cite{Weller2005, Emery2005} 
The temperature dependence of the magnetic penetration depth $\lambda_{ab}(T)$ shows an exponential behavior at low temperatures, characteristic for a $s$-wave2 pairing mechanism. \cite{Lamura2006} 
The temperature and magnetic field dependences of the electronic specific heat are consistent with a fully gapped superconducting order parameter. \cite{Kim2006heat}
Isotope effect studies show that superconductivity in CaC$_6$ is dominated by coupling of the electrons by Ca phonon modes. \cite{Hinks2007}
It has been argued that coupling with low energy Ca vibrations must be very strong to obtain the observed $T_{c}$, in disagreement with the specific heat data that suggests a weak coupling regime. \cite{Mazin2007}
For an overview of theoretical and experimental studies see ref. \onlinecite{Kim2007}.
Despite of the recent efforts the origin of the superconducting mechanism in CaC$_6$ is still not fully understood.

One of the important parameters of a superconductor is the magnitude and the temperature dependence of the superconducting gap size.
The gap $\Delta_{0}$ of CaC$_6$ was measured \cite{Bergeal2006} by scanning tunneling spectroscopy, $2\Delta_{0} = 25.6 \pm 3.2 $\,cm$^{-1}$ yielding $2\Delta_{0} / 2 k_{B} T_{c} = 3.35 \pm 0.42 $, slightly less than the weak coupling BCS value of 3.53. 
A subsequent scanning tunneling spectroscopy experiment\cite{kurter2007} reported a much larger gap size  $2\Delta_{0} = 37.1 \pm 3.2$ \,cm$^{-1}$, with $2\Delta_{0} / 2 k_{B} T_{c} \approx 4.6$ in the strong coupling regime. 
This discrepancy of 50\% in the gap values deserves further clarification.
The temperature dependence and the magnitude can be easily tested by infrared spectroscopy. \cite{Glover1956}
Infrared photons with energy larger than $2\Delta_{0}$  break a Cooper pair in a process of scattering off impurities and create a pair of quasiparticles. \cite{Mattis1958,Popel1991}
The signature of the superconducting gap in infrared reflectance spectra increases along with the scattering rate of quasiparticles. Here we report the results of an infrared reflectance spectroscopy experiments on CaC$_6$ which reveal a distribution of superconducting gaps with the average value $2\Delta_{0} / 2 k_{B} T_{c}$ consistent with the weak coupling BCS value of 3.53. 
The gap distribution is in very good agreement with the distribution obtained from an \textit{ab initio} density functional calculations by Sanna \textit{et al.} \cite{Sanna2007}

In our study we investigated the reflectance $R$ of single crystals of CaC$_6$ between 3\,K and 15\,K. 
We found typical signatures of the superconducting gap in the reflectance ratios $R_{sc}/R_{n}$ of the superconducting and the normal state, respectively, and followed its temperature dependence. 
The appearance of the gap signature in $R_{sc}/R_{n}$ indicates that CaC$_6$ is in the dirty limit with the quasiparticle scattering rate large compared to $2\Delta_{0}$.


We measured the reflectivity of total of three samples, 26, 31, and 42, grown in Stuttgart \cite{Kim2006heat}, with dimensions in the $(ab)$ plane  $3.7 \times 3.5 \, \mathrm{mm^{2}}$, $5.3 \times 3 \, \mathrm{mm^{2}}$, and $3.5 \times 1.2 \, \mathrm{mm^{2}}$, respectively. 
The sample 31 was a mosaic composed of two pieces.
The surfaces used for the reflectance measurements on samples 31 and 42 were obtained by cleaving the crystals while for 26 the outer surface of an as-grown sample was investigated.
The superconducting properties of the samples were characterized by magnetic susceptibility and d.c. measurements.
The conductivity $\sigma_{dc}$  measured in the normal state at 12\,K was $1.3 \, (\mu \Omega \, \mathrm{cm})^{-1}$.
Samples were loaded in a glove box in a controlled argon atmosphere into the reflectance probe. 
After mounting into the cryostat Ar gas was replaced by He heat exchange gas.
A polarizing Martin-Puplett interferometer SPS200 (Sciencetech Inc.) equipped with a $^3\mathrm{He}$-cooled bolometer kept at $0.3 \,\mathrm{K}$ was used to measure the spectra between  4 and $80\,\mathrm{cm^{-1}}$.
A mercury arc lamp was used as a light source.
A small wire-grid polarizer was placed at the exit port of a light pipe inside the sample chamber in front of an aluminum-coated glass mirror which focused light on the sample. 
The $\mathbf{E}$ vector of light was perpendicular to the plane of incidence and parallel to the plane of carbon atoms, the $(ab)$ plane.
With this scattering geometry the conductivity in the $(ab)$ plane can be probed without contaminations from the $c$-axis conductivity.
Light intensity reflected by the sample  at $T < T_{c}$ was divided by the light intensity reflected by the sample in the normal state at 15\,K.

\begin{figure}[dispersions]
\includegraphics[width=8.6cm]{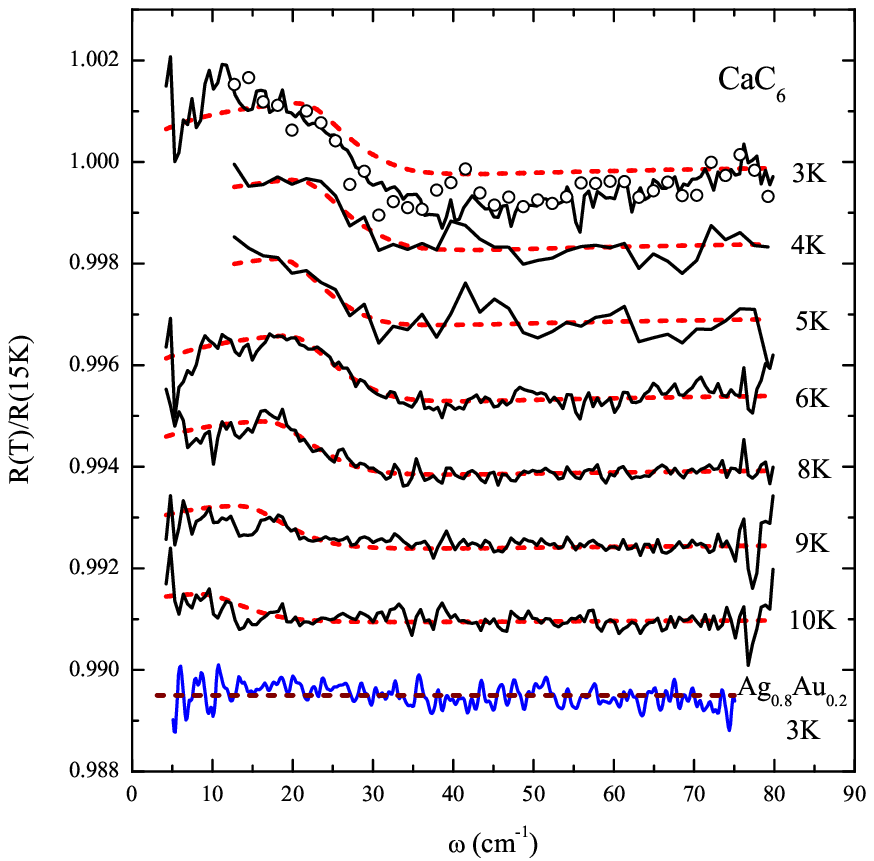}
\caption{
Reflectance spectra of CaC$_6$ (samples 31 and 42) in the superconducting state relative to the normal-conducting state at 15\,K.
Starting from 4\,K the spectra are offset by $-0.0015$ in vertical direction. 
4\,K and 5\,K spectra were measured only on sample 42, the smaller size of which limited the spectral range and signal to noise ratio.
Open circles represent the 3\,K spectrum of sample 42.
Dashed lines are fit results assuming a BCS superconductor with a scaled gap distribution\cite{Sanna2007} below $T_c$ and Drude-like conductivity at 15\,K. 
}
\label{s31gapDwaterfall}
\end{figure}

The $T$ dependence of relative reflectance is shown in Fig.\ref{s31gapDwaterfall}.
A step-like  feature observed  below 30\,cm$^{-1}$ with its largest amplitude at 3\,K decreases and shifts to lower frequency as $T$ approaches $T_c$.
To rule out the possibility of the step-like feature in the CaC$_6$ spectra being an artefact of the measurement method, a reference sample made of solid solution Ag$_{0.8}$Au$_{0.2}$ was measured simultaneously with the CaC$_6$ samples.
Below 15\,K, Ag$_{0.8}$Au$_{0.2}$  has a very small $T$ dependence of the resistivity. \cite{Fujishiro1994}
Between between 4 and 80\, cm$^{-1}$ the  reflectance ratio $R(3\mathrm{K})/R(15\mathrm{K})$ obtained from  Ag$_{0.8}$Au$_{0.2}$ was a flat line with a noise to signal amplitude $\pm 2.5\times10^{-4}$.
The deviation from the flat line is an order of magnitude smaller than the step size features observed in $R(T)/R(15\mathrm{K})$ spectra of CaC$_6$ (Fig.\ref{s31gapDwaterfall}).

To fit the data we calculated the zero angle reflectance using well-known relations between the complex dielectric constant  $\epsilon(\omega)$ and the complex conductivity $\sigma(\omega)$ assuming a Drude-like conductivity for the normal state. \cite{Dressel2002Book}
In the superconducting state the expression by Zimmermann \textit{et al.}\cite{Zimmermann1991} for a BCS superconductor was used.
The Zimmermann expression yields the complex conductivity normalized to the normal state conductivity $\sigma_{dc}$ and depends on parameters $\Delta_{0}$,  $T_c$,  $T$, and the scattering rate $\gamma$.
To calculate the (relative) reflectance, the d.c. conductivity $\sigma_{dc}$ and the dielectric constant $\epsilon_{\infty}$ are needed.
We used $\epsilon_{\infty}=4$, where contributions of all lattice and electronic oscillators above 80\,cm$^{-1}$ are included.
$\sigma_{dc}$ is related to the plasma frequency and the scattering rate according to $\omega_p^2 =60\gamma\sigma_{dc}$, and  one of them, $\sigma_{dc}$ or $\omega_p$ is needed as an additional input parameter.
The expression of Zimmermann  \textit{et al.} can be extended to superconductors with a distribution of $\Delta$ by assuming that the conductivity in the superconducting state is given by the superposition of conductivities according to 
$\sigma_{sc}(\omega) = \sum_{i}  f (2 \Delta _{i}) \sigma_{sc}(\omega, 2 \Delta _{i}) \, / \, \sum_{i}f(2 \Delta _{i})$,
where  $f(2 \Delta _{i})$ is the distribution function of the gap.
In the fitting procedure of $R(T)/R(15\mathrm{K})$ the parameters $\omega_p$,  $T_c$ and  $T$ were fixed.
In the first step a temperature independent value for $\sigma_{dc}$ was obtained for each sample and used in the second step to perform a complete fit of $\Delta(T)$.
In the case of gap distribution $f(s(T)\cdot 2 \Delta_{i})$ the result was the scaling factor $s(T)$, where    $s(0)=1$.
To compensate for detector drifts an additional fitting parameter $k$ was introduced which multiplied the relative reflectance and ranged from 0.9982 to  0.9999.

\begin{figure}[dispersions]
\includegraphics[width=8.6cm]{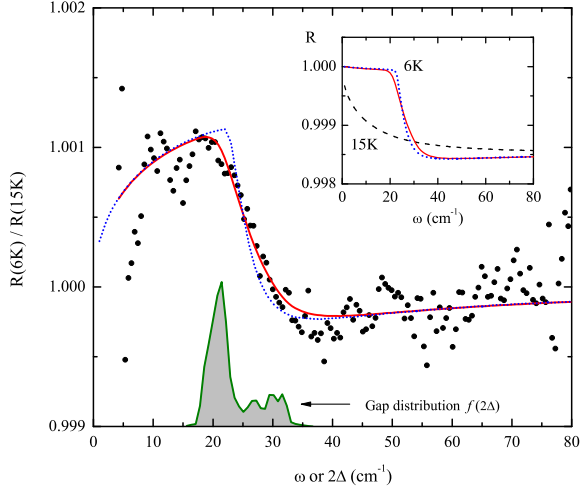}
\caption{
Reflectance of CaC$_6$ sample 31 at 6K relative to 15K (dots). 
Solid line is best fit using the gap distribution\cite{Sanna2007} scaled by a factor $s \, (6\,\mathrm{K})=0.94$ (shaded area). 
The dotted line was obtained assuming an isotropic BCS superconductor with a  gap  $2 \Delta\,(6\,\mathrm{K}) = 22.6$ \,cm$^{-1}$. 
Inset: the dashed line is calculated absolute reflectance in the normal conducting state; solid and dotted lines are the calculated absolute reflectances at 6\,K for the gap distribution and single gap, respectively, as described in the text.
}
\label{fig_s31gapD6K}
\end{figure}

\begin{figure}[dispersions]
\includegraphics[width=8.6cm]{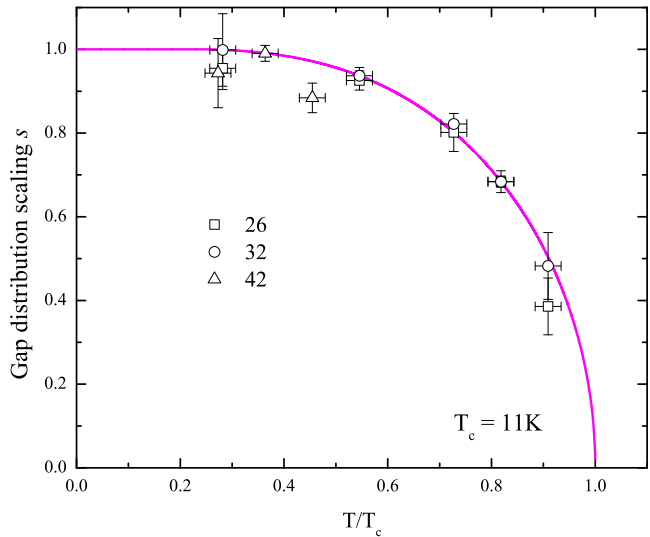}
\caption{
Temperature dependence of the gap distribution scaling factor $s(T)$ from fitting relative reflectance spectra for samples 26, 31, and 42. 
The solid line is the temperature dependence of the gap of a BCS superconductor.
Vertical error bars are three times the standard error as calculated from the fit covariance matrix. 
The temperature error is estimated to be about 2.5\%.
}
\label{fig_TwoDeltaTdep}
\end{figure}

The fits assuming two different gap models and corresponding absolute reflectances are shown in Fig.\ref{fig_s31gapD6K}.
The absolute reflectance in the superconducting state at frequencies below  $2 \Delta$ is unity and drops at frequencies above the gap,  $\omega > 2 \Delta$ (see inset to Fig.\ref{fig_s31gapD6K}).  
There are two different regimes in the normal conducting state. \cite{Dressel2002Book} 
At low frequencies $\omega \ll \gamma$, the Hagen-Rubens square root behavior is observed, $R \approx 1 - (2 \omega  /15\sigma_{dc})^{1/2} $.
At higher frequencies $\omega \gg \gamma$, the reflectance flattens to $R \approx   1 - (\gamma / 15\sigma_{dc})^{1/2}$.
First we fitted the 6\,K spectrum using the single gap model and the plasma frequency $\omega_{ab} = 6.6 \, eV \approx 53000 \,  \mathrm{cm}^{-1}$ as calculated by Mazin \textit{et al.} \cite{Mazin2007}
The fit  obtained with $2\Delta (6K) = 22.6$ \,cm$^{-1}$ is not perfect since the measured spectrum lacks a sharp onset of full reflectivity at $2\Delta$.
The fit can be significantly improved (\textit{cf.} Fig.\ref{fig_s31gapD6K}) if the gap distribution as obtained from the first principles density functional calculation of Sanna \textit{et al.} \cite{Sanna2007} is used. \cite{rigorous}
The gap model of Sanna  \textit{et al.} scaled by a factor of $s(T) < 1$ was used to to fit  the whole temperature range of relative reflectances (\textit{cf.} Fig.\ref{s31gapDwaterfall}).
The fit of relative reflectance spectra yielded normal state conductivities $\sigma_{dc} = 0.6 \, (\mu \Omega \, \mathrm{cm})^{-1}$ for sample 26 and  $1.2 \, (\mu \Omega \, \mathrm{cm})^{-1}$ for samples 31 and 42, in good agreement with the measured d.c. conductivity (see above).
Although the fit of relative reflectance gave a lower conductivity for sample 26, the quality of the fit (not shown) and the scaling factor $s(T)$ were similar to that of other two samples (\textit{cf.} Fig.\ref{fig_TwoDeltaTdep}).
Apparently there is a higher level of defects in the as-grown surface of sample 26 that reduce the conductivity but do not affect the superconducting state. 

In the following we discuss the results obtained for the samples with the higher conductivity if not otherwise stated. 
The scattering rate obtained, $\gamma = \omega_{ab}^{2}\, / \,60\, \sigma_{dc} =39$ \,cm$^{-1}$, is not much larger than the maximum value of twice the gap.
Assuming a Fermi velocity  $v_{F} = 5.3 \times 10^{7} $\,cm\,s$^{-1}$ (Ref. \onlinecite{Mazin2007}) the calculated  mean free path amounts to  $l_{mfp} = v_{F} / \gamma = 450$\,nm.
The coherence length  determined by other experimental probes \cite{Emery2005,Bergeal2006,Kim2006heat,Jobiliong2007} is rather small, $\xi_{ab} = 33$\,nm, compared to the mean free path $l_{mfp}$.
From  $l_{mfp} \gg \xi_{ab}$ we must conclude that CaC$_6$ is in the clean limit. 
On the other hand the maximum value of the gap in the distribution function, $2 \Delta_{max} \approx 35$\,cm$^{-1}$, puts CaC$_6$ into intermediate limit, between dirty ($ 2\Delta_{0}<\gamma $) and clean ($ 2\Delta_{0}>\gamma $).
Lamura \textit{et al.} \cite{Lamura2006} concluded that $\mathrm{CaC}_{6}$ is in the dirty limit.
Also conduction electron spin resonance measurements suggest that superconductivity in CaC$_6$ can be described in the dirty limit. \cite{muranyi2008}
This contradiction,  $l_{mfp} \gg \xi_{ab}$ (clean limit)  versus $2 \Delta_{max} \approx \gamma$ (intermediate limit), could be due to the overestimation of $l_{mfp}$. 
The mean free path $l_{mfp}$ is calculated from $\sigma_{dc}$ using theoretical estimates for plasma frequency and Fermi velocity and will be reduced by increasing $\omega_p$ and  reducing $v_F$.
Indeed,  from the comparison of  $\xi_{ab}=33$\,nm to a theoretical value of coherence length\cite{Tinkham1996} $ \xi_0=\hbar v_F/\pi \Delta_{0}=80$\,nm (we take $\Delta_{0}=11$\,cm$^{-1}$ as the maximum of gap distribution) we see that the theoretical value of $v_F$ is overestimated.

The gap scaling factor $s(T)$ (Fig. \ref{fig_TwoDeltaTdep}) follows the temperature dependence of a $s$-wave superconductor.\cite{Muhlschlegel1959}
The experimental errors close to $T_{c}$, however, are too large to obtain an independent determination of $T_{c}$ from  reflectance spectra.
Smaller than expected reflectance above the gap in $3 \,$K spectra (Fig.\ref{s31gapDwaterfall}) can be caused by strong coupling effects that reduce the imaginary part of the complex conductivity. \cite{Shaw1968}
It is not clear why such a deviation appears (for all measured samples) only at 3\,K  and not at 4\,K, both temperatures substantially less than $T_c$. 
We saw no evidence of the expected low-frequency $E_{u}$ phonon mode \cite{Calandra2005, Calandra2006, Kim2006pressure, Hlinka2007} around $115 $\,cm$^{-1}$ as it was outside the limits of optimal performance of our measurement set-up.

To conclude, we have measured the temperature dependence of far infrared reflectance spectra of  CaC$_6$.
The results suggest that CaC$_6$ is a $s$-wave superconductor with an anisotropic gap.
Our results provide evidence of strong coupling effects in the spectra at 3\,K.
The gap signature in the far infrared spectra indicates that CaC$_6$ is in the dirty limit, in accordance with other experiments.

\section{Acknowledgments}
Support by Estonian Science Foundation grants 5553, 6138, and 7011 is acknowledged.  

\bibliographystyle{apsrev}


\end{document}